\documentclass{article}
\usepackage[a4paper]{geometry}
\usepackage{epsf}
\usepackage{epsfig}
\usepackage{amsmath}
\usepackage{amssymb}

\begin{document}
\begin{center}
{\Large \bf  Two-loop {\boldmath ${\cal O}(\alpha_s y^2 )$
and ${\cal O}(y^4)$} MSSM corrections to the 
pole mass of the $b$-quark}
\\ \vspace*{5mm} A.~Bednyakov, A.~Sheplyakov 
\end{center}

\begin{center}
Bogoliubov Laboratory of Theoretical Physics, \\
Joint Institute for Nuclear Research, Dubna, Russia
\end{center}

\begin{abstract}
The  paper is devoted to the calculation of the  two-loop  
${\mathcal O}(\alpha_s y^2)$ and ${\mathcal O}(y^4)$ MSSM
corrections to the relation between the pole mass of 
the $b$-quark and its running mass in $\overline{\rm DR}$
scheme.  To evaluate the needed diagrams the large
mass expansion procedure is used. The obtained contributions
are negative in most of the regions of the parameter space
and partly compensate the positive 
${\mathcal O}(\alpha_s^2)$ contribution calculated earlier.
\end{abstract}

\begin{section}{Introduction}
\label{intro}

The Minimal Supersymmetric Standard Model (MSSM) became an underlying
framework for both the theoretical and experimental research in
supersymmetric phenomenology.  During recent years considerable
progress has been achieved in calculation of various processes
involving supersymmetric particles and the parameter space of
the MSSM is severely constrained by numerous phenomenological
fits \cite{Ellis:2004bx,deBoer:2003xm,Baer:2003wx,deBoer:2001xp,Abel:2000vs}.  
These fits require precision calculations of radiative corrections
due to virtual supersymmetric particles. Because of complexity of
the MSSM Lagrangian these calculations are very lengthy and assume
the creation of the proper computer codes.

In this paper we calculate two-loop radiative corrections to
the pole mass of the $b$-quark within the MSSM.  The one-loop
correction has been calculated in \cite{Pierce:1996zz} and is
very essential. It reduces considerably the value of the running
quark mass extracted from the pole mass.  It should be noted that
the leading ${\mathcal O}(\alpha_s)$ contribution from stop-gluino
loop which is positive  is reduced significantly by large negative
correction from stop-chargino loop that is proportional to the Yukawa
couplings of heavy quarks ($y_t, y_b$), $\tan\beta$ and the Higgs
mixing parameter $\mu$. For large $\tan\beta$ one has to take into
account not only $y_t$ but also $y_b$ since it becomes large as well.
Calculation of the two-loop correction of the order of ${\mathcal
O}(\alpha_s^2)$ has been performed in \cite{Bednyakov:2002sf}.
It was found that this correction is also positive and of the same
order of magnitude as the one-loop MSSM contribution.

The aim of this paper is to present our calculations of the 
corrections proportional to the Yukawa couplings. We calculate
the two-loop ${\mathcal O}(\alpha_s y^2)$ and ${\mathcal O}(y^4)$
MSSM corrections to the relation between the pole mass of the
$b$-quark and its running mass in $\overline{\rm DR}$ scheme
\cite{Siegel:1979wq}.
The $\overline{\rm DR}$ scheme corresponds to a theory regularized
by dimensional reduction (from space-time dimension $d=4$ to 
$d=4-2\varepsilon$) and renormalized minimally. This allows one
to keep supersymmetry unbroken and use the computational advantages
of dimensional regularization.

To evaluate the needed diagrams we made use of the large mass
expansion procedure  \cite{Smirnov:2002pj}
and restrict ourselves to the terms up to
${\mathcal O}(m_b^2/M_{hard}^2)$, where $M_{hard}$ stands for
all mass scales involved in the problem that are much larger than
$m_b$. Since the total number of diagrams exceeds 1000, we are not
able to present the calculation in a compact form, so we provide
numerical demonstration of the results. The final result is also
available upon request in a form of C++ code\footnote{E-mail:
mailto:varg@thsun1.jinr.ru.}.
As expected the calculated corrections are negative in most of the
regions of the parameter space of the MSSM and partly compensate
those proportional to ${\cal O}(\alpha_s^2)$. By absolute value
the ``leading'' ${\mathcal O}(\alpha_s^2 + \alpha_s y^2 + y^4)$
two-loop corrections are of the order of 30 to 40 percent of the
one-loop ones.
\end{section}
	
\begin{section}{Pole mass of $b$-quark}
\label{polemassb}

The pole mass of a particle is defined as the real part of complex
pole of resumed propagator  (we discuss only perturbative effects).
Full connected propagator of a quark can be written
as\footnote{The CKM matrix is supposed to be diagonal.}
\begin{equation}
\label{fullprop}
\frac{i}{\hat{p} - m - \Sigma(\hat{p}, m_{i})},
\end{equation}
where
\begin{equation}
\Sigma(\hat{p}, m_i) = \hat{p} \Sigma_V(p^2, m_i^2) 
	+ \hat{p} \gamma_5 \Sigma_A(p^2, m_i^2)
	+ m \Sigma_S(p^2, m_i)
\end{equation}
is the self-energy of the quark, so the pole mass $M_{pole}$
satisfies the following equation
\begin{equation}
\label{polemassdef}
\left(1+\Sigma_V(M_{pole}^2, m_i^2)\right)^2 M_{pole}^2 
	- \Sigma_A^2(M_{pole}^2, m_i^2) M_{pole}^2  
	- m^2 \left(1-\Sigma_S(M_{pole}^2, m_i)\right)^2 = 0.
\end{equation}
Solving this equation perturbatively, one gets
\begin{eqnarray}
\label{pole2bare}
&& \frac{M_{pole}-m}{m} = \alpha M^{(1)} + \alpha^2 M^{(2)}, \quad \mbox{where} \\
&& M^{(1)} = \Sigma_V^{(1)}(m^2, m_i^2) + \Sigma_S^{(1)}(m^2, m_i), \\
&& M^{(2)} = \Sigma_V^{(2)}(m^2, m_i^2) + \Sigma_S^{(2)}(m^2, m_i) +
\frac{1}{2} {\Sigma_A^{(1)}}^2(m^2, m_i^2) \nonumber\\
&& + M^{(1)} \left( \Sigma_V^{(1)}(m^2, m_i^2) 
	+ 2 m^2 \frac{\partial}{\partial p^2}
	\left( \Sigma_V^{(1)}(p^2, m_i^2) + \Sigma_S^{(1)}(p^2, m_i) \right)|_{p^2=m^2}\right) \;,
\end{eqnarray}
and $\alpha$ stands for all couplings of the theory.

Using eq. \eqref{pole2bare}, one calculate relation between pole
and running masses of the $b$-quark (depending on the prescription
used for evaluation of $\Sigma^{(i)}$, mass parameter $m$ in
(\ref{pole2bare}) may correspond to bare or renormalized mass,
so one obtains relation between pole and bare masses or between
pole and running masses).

We use regularization by dimensional reduction $\overline{\rm DR}$,
a modification of the conventional dimensional regularization,
originally proposed in \cite{Siegel:1979wq} and the same renormalization
prescription as in \cite{Bednyakov:2002sf, Avdeev:1997sz}.
Therefore we calculate
\begin{equation}
\frac{\Delta m_b}{m_b} \equiv \frac{M_b^{pole} - m_b^{run}}{m_b^{run}},
\label{pole2DR}
\end{equation}
where $M_b^{pole}$ stands for the pole mass of the $b$-quark and
\mbox{$m_b^{run}\equiv m_b(\bar{\mu})$} corresponds to the running
\mbox{$\overline{\rm DR}$ mass} of the $b$-quark at the scale 
$\bar{\mu}$.

To evaluate this quantity, we have to calculate more than 1000
two-loop propagator type diagrams (see Fig.~\ref{twoloopdiag}). In
order to simplify our calculations, we used the t'Hooft-Feynman
gauge ($\xi_W=1, \xi_Z=1$). We also neglected mixing in the chargino
and neutralino sectors, so only higgsino states were taken into
account. This implies that all higgsinos have the same mass
which is equal to Higgs mixing parameter $\mu$.
\begin{figure}[t]
\hspace*{-0.6cm}
\includegraphics[scale=0.60]{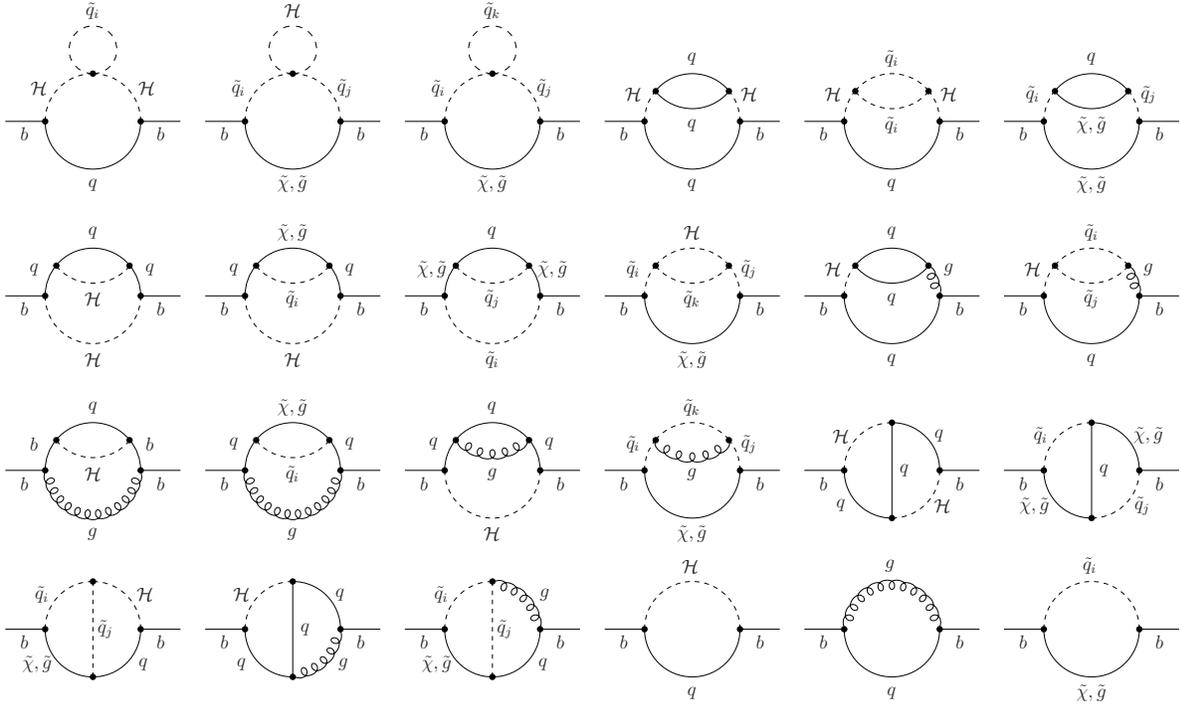}
\caption{
	Feynman diagrams contributing to quark self-energies,
	Here $q$, $\tilde q_i$, $g$, $\tilde g$ denote quark, squark,
	gluon and gluino fields respectively,
	${\cal H}=\{h_0, H_0, A_0, G_0, H^+, G^+\}$ corresponds
	to various Higgs fields and 
	$\tilde \chi = \{\tilde \chi_0, \tilde \chi^+\}$ are
	higgsino states.
	Two-loop diagrams of the order 
	${\cal O}(\alpha_s^2)$ are not presented here (see, e.g \cite{Bednyakov:2002sf}).
	}
\label{twoloopdiag}
\end{figure}

Due to presence of many different mass scales, the problem of exact
evaluation of two-loop diagrams is rather complex.  Exploiting the
fact that
\begin{equation}
\label{masshier}
m_b \ll m_t, m_{h_0}, m_{H_0}, m_{H^+}, m_{\tilde{q}},
m_{{\tilde\chi}^0}, m_{{\tilde\chi}^+}, m_{G_0}, m_{G^+},
\end{equation} 
where $m_b$ and $m_t$ are masses of bottom and top quarks,
respectively, $m_{h_0}, m_{H_0}, m_{H^+}$ are masses of the
Higgs bosons, $m_{{\tilde\chi}^0}$ is the neutralino mass,
$m_{{\tilde\chi}^+}$ is the chargino mass, $m_{\tilde{q}}$ are
masses of different squarks, $m_{G_0}, m_{G^+}$ are masses of
pseudo-goldstone bosons,
we use the method of large mass expansion \cite{Smirnov:2002pj}
to reduce the evaluation of multi-scale two-loop diagrams to the
calculation of two-loop vacuum integrals and products of one-loop
on-shell propagator-type diagrams and one-loop bubble integrals.
We include only terms up to $\mathcal{O}(m_b^2/M_{hard}^2)$, where
$M_{hard}$ denotes any mass of the right hand side of \eqref{masshier}.

A general Feynman diagram $F_\Gamma$ which depends on the large
masses $M_1, M_2, \ldots,$ small masses $m_1, m_2, \ldots$ and
small external momenta $p_1, p_2, \ldots$ can be expanded as follows
\cite{Smirnov:2002pj}:
\begin{equation}
\label{l:asympexp}
F_\Gamma(p_1, \ldots, M_1, \ldots, m_1, \ldots) =
\sum_\gamma F_{\Gamma/\gamma}(p, m) \mathcal{M}_\gamma(p_\gamma, m)
F_\gamma(M, m, p_\gamma),
\end{equation}
where operator $\mathcal{M}_\gamma(p_\gamma, m)$ performs Taylor
expansion in small external (with respect to subgraph~$\gamma$)
momenta and masses. The sum runs over all asymptotically irreducible
subgraphs of original graph $\Gamma$.

For calculation of a two-loop propagator diagram that does not
have cuts composed of two light lines up to $\mathcal{O}(p^2/M^2)$
one can use naive expansion, when sum in (\ref{l:asympexp}) runs over
only the trivial subgraph~$\gamma=\Gamma$ (see Fig.~\ref{naiveExp}).
\begin{figure}
\begin{center}
\mbox{\includegraphics[scale=1]{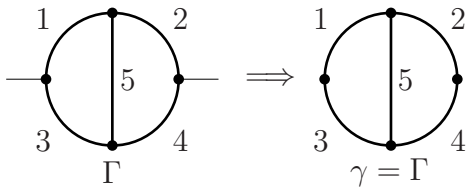}}
\caption{The diagram has no cuts composed of two light lines, therefore
the naive expansion can be used.}
\label{naiveExp}
\end{center}
\end{figure}
In case of a diagram that has such a cut a non-trivial subgraph
$\gamma_1$ has to be taken into account (see Fig.~\ref{subgExp}).
\begin{figure}
\begin{center}
\includegraphics[scale=1]{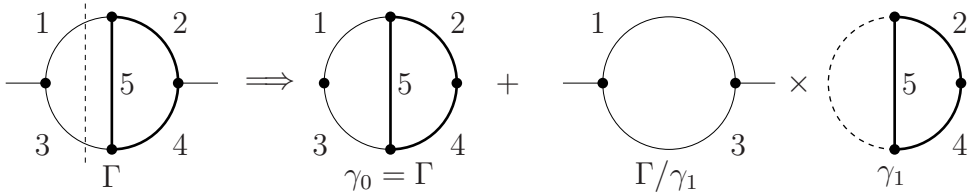}
\caption{The diagram has one cut composed of two light lines, so
non-trivial subgraph $\gamma_1$ has to be taken
into account.}
\label{subgExp}
\end{center}
\end{figure}

Two-loop vacuum integrals can be recursively reduced to a
master-integral \cite{Davydychev:1992mt} by integration by
parts method \cite{Chetyrkin:1981qh}.

Our calculation can be performed in a semi-automatic way.
First, FeynArts \cite{Hahn:2001rv} is used to generate
the diagrams. Then the contribution of individual diagrams to
the relation \eqref{pole2bare} is evaluated by means of C++
program TwoLoop, based on GiNaC \cite{Bauer:2000cp}.  TwoLoop
\cite{twoloopvargs} performs large mass expansion according to
\eqref{l:asympexp}, uses recurrence relations of \cite{Davydychev:1992mt}
to reduce two-loop vacuum integrals to the master integral, and
performs expansion in $\varepsilon$.
Results of this calculation were analytically cross-checked using
FORM program based on ON-SHELL2 package \cite{Fleischer:1999tu}.

After evaluation of two-loop diagrams one should substitute them into
\eqref{pole2bare} and perform renormalization in order to express
the correction \eqref{pole2DR} in terms of running $\overline{\rm
DR}$ parameters.  Counter-terms were calculated and checked in
the same way as in \cite{Bednyakov:2002sf}, e.g. our renormalization
constants for the gauge and Yukawa couplings reproduce well-known
$\beta$-functions \cite{Ibanez:1984vq}.
\end{section}

\begin{center}
\begin{figure}
\includegraphics{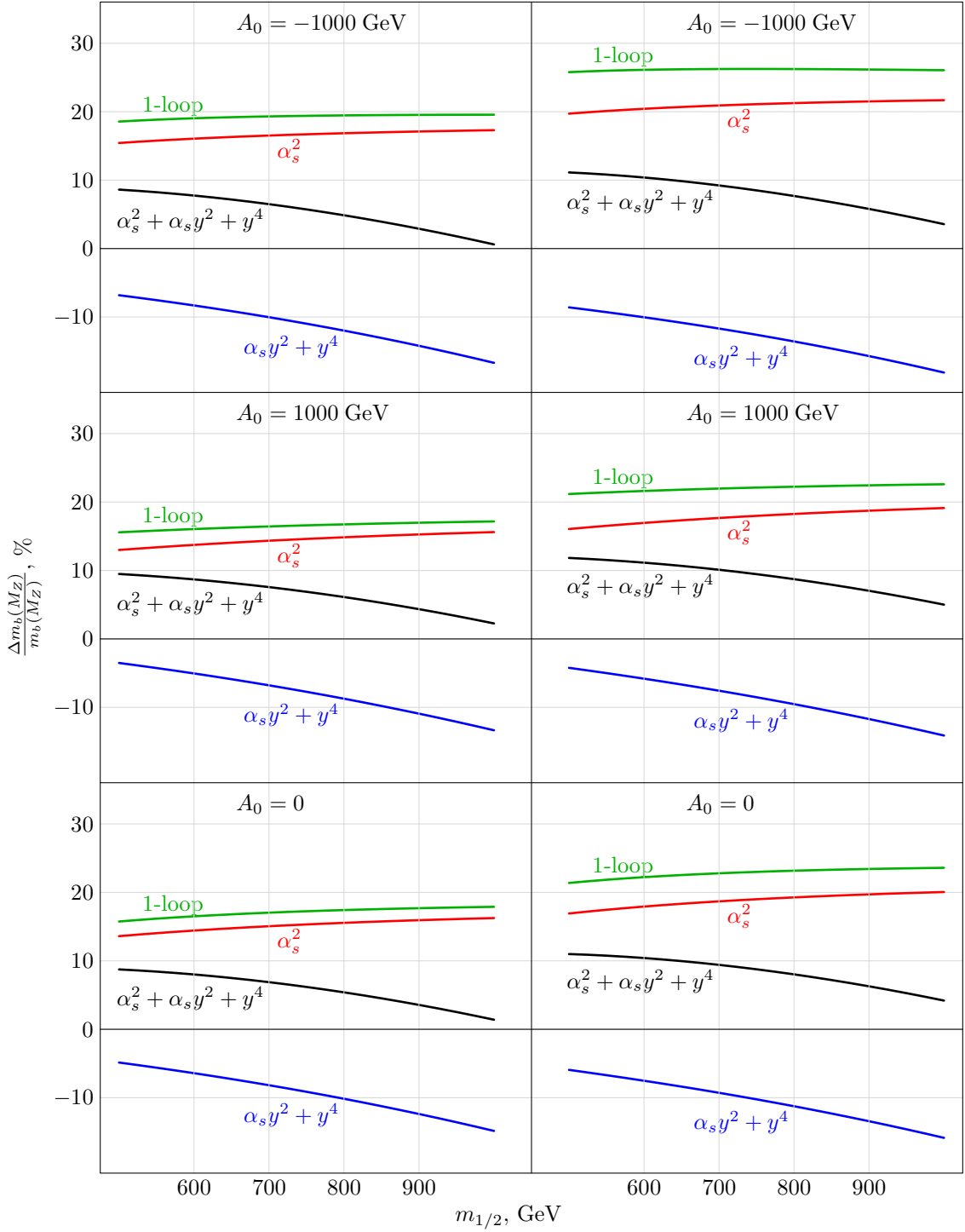}
\caption{Different MSSM contributions to $\Delta m_b(M_Z)/m_b(M_Z)$.
Here $m_0 = 1000 \; {\rm GeV}$, left column corresponds to 
$\tan\beta = 35$, right column corresponds to $\tan\beta = 50$.}
\label{m12plots}
\end{figure}
\end{center}

\begin{figure}
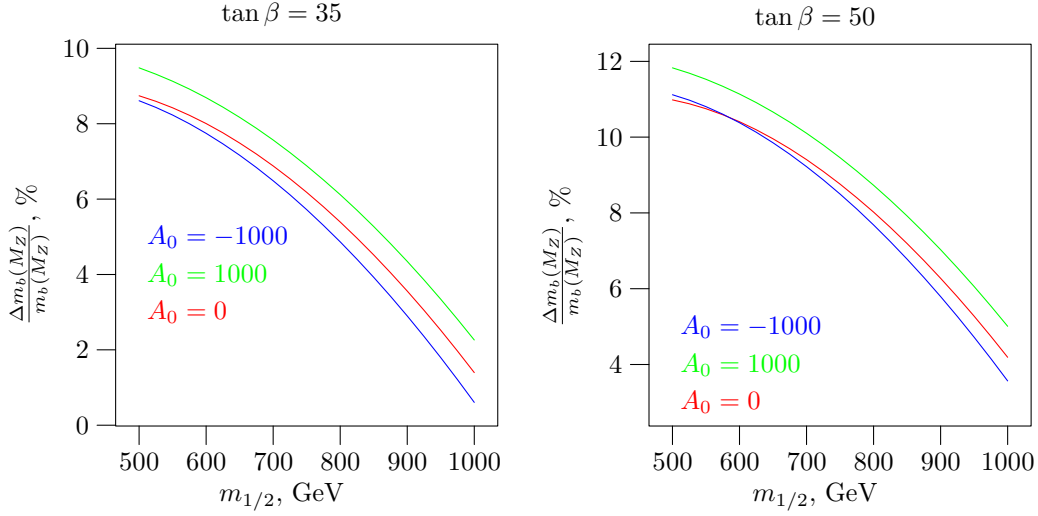

\begin{tabular}{cc}
 \includegraphics{plots-2} & 
 \includegraphics{plots-3}
\end{tabular}
\caption{ $A_0$ dependence of ${\mathcal O}( \alpha_s^2 + y^4 + \alpha_s y^2)$ 
two-loop correction to $\Delta m_b(M_Z)/m_b(M_Z)$. 
Here $m_0 = 1000 \; {\rm GeV}$.
Varying $A_0$ from $-1000 \; {\rm GeV}$ to $1000 \;
{\rm GeV}$ does not change behaviour of the correction significantly.}
\label{AzeroDeps}
\end{figure}

\begin{figure}
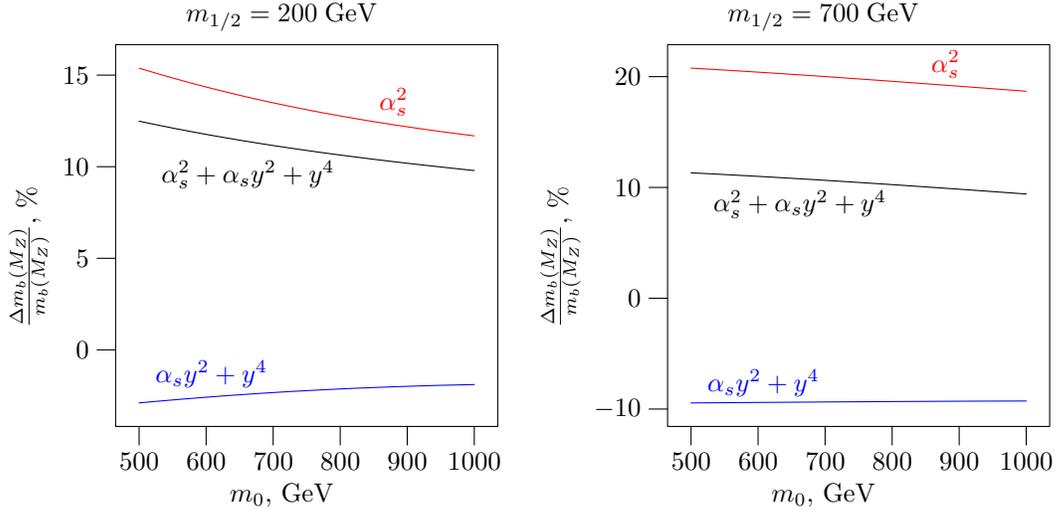

\begin{tabular}{cc}
 \includegraphics{plots-4} &
 \includegraphics{plots-5}
\end{tabular}
\caption{ Different two-loop MSSM corrections
to $\Delta m_b(M_Z)/m_b(M_Z)$ as a function of $m_0$. 
Here $m_0 = 1000 \; {\rm GeV}, A_0 = 0, \tan\beta = 50$.
Dependence of ${\mathcal O}(\alpha_s y^2 + y^4)$ correction on
$m_0$ is essential for $m_{1/2} = 200 \; {\rm GeV}$.
On the other hand, for $m_{1/2} = 700 \; {\rm GeV}$ the change 
of $m_0$ from $400 \; {\rm GeV}$ to $1000 \; {\rm GeV}$
yields change of ${\mathcal O}(\alpha_s y^2 + y^4)$ correction 
less than 10\%.}
\label{mZeroDeps}
\end{figure}

\begin{section}{Numerical results}
\label{numres}
The analytical result of our calculation is very complicated due
to presence of a large number of masses, and no phenomenologically
acceptable limit seems to exist.  Therefore we present here the
numerical analysis of our results.

While a detailed scan over the more-than-hundred-dimensional
parameter space of MSSM is clearly not practicable, even a sampling
of the three- (four-)dimensional CMSSM parameter space of $m_0$,
$m_{1/2}$ and $A_0$ ($\tan \beta$) is beyond the present capabilities
of phenomenological studies, especially when one tries to simulate
experimental signatures of supersymmetric particles within a detector.
For this reason, one often resorts to specific benchmark scenarios,
i.e., one studies only specific parameter points or at best samples
the one-dimensional parameter space \cite{modelline, Martin:2002zb}

Numerical values of running SUSY parameters at the $M_Z$ scale have
been calculated as a function of CMSSM parameters  with the program
TwoLoop \cite{twoloopvargs} after interfacing it to a slightly
modified version of SOFTSUSY \cite{Allanach:2001kg} in the framework of
mSUGRA supersymmetry breaking scenario. We only consider $\mu > 0$
and large values of $\tan\beta$, since small $\tan\beta$ and negative
$\mu$ seem to be excluded by experimental data \cite{deBoer:2003xm}.

Fig.~\ref{m12plots} shows $m_{1/2}$-dependence of ${\mathcal
O}(\alpha_s^2)$ \cite{Bednyakov:2002sf} and ${\mathcal O}(\alpha_s
y^2 + y^4)$ MSSM contributions to \eqref{pole2DR}. For comparison,
full one-loop MSSM correction was plotted. Numerical value of the
${\mathcal O}(\alpha_s^2)$ contribution is comparable with the full
one-loop result, so neglecting other two-loop MSSM corrections leads
to a contradiction with the perturbation theory. The ${\mathcal
O}(\alpha_s y^2 + y^4)$ contribution is negative in a wide range of
the CMSSM parameter space and reduces ${\mathcal O}(\alpha_s^2)$
contribution by approximately $30 - 80 \%$. As a consequence
${\mathcal O}(\alpha_s^2+\alpha_s y^2 + y^4)$ two-loop correction
appears to be less than 40\% of the one-loop contribution and
therefore can be used in phenomenological studies of CMSSM.

Varying $A_0$ in a wide range from $-1000 \; {\rm GeV}$ to $1000 \;
{\rm GeV}$ does not change the behaviour of ${\mathcal O}(\alpha_s^2
+ \alpha_s y^2 + y^4)$ correction significantly, as one can see
from Fig.~\ref{m12plots}.  To make this fact more clear, several
curves corresponding to different values of $A_0$ were plotted
in Fig.~\ref{AzeroDeps}.

In Fig.~\ref{mZeroDeps} we present dependence of different two-loop
MSSM contributions on $m_0$. For $m_{1/2} \gtrsim 500 \; {\rm GeV}$
the change of $m_0$ from $400 \; {\rm GeV}$ to $1000 \; {\rm GeV}$
yields change of ${\mathcal O}(\alpha_s y^2 + y^4)$ correction less
than 10\%. On the other hand, dependence of ${\mathcal O}(\alpha_s y^2 +
y^4)$ correction  on $m_0$ becomes essential for relatively small
$m_{1/2} \; ( m_{1/2} \approx 200 \; {\rm GeV})$.

It should be noted, that in considered range of CMSSM parameters
space ${\mathcal O}(\alpha_s^2 + \alpha_s y^2 + y^4)$ correction
is positive.
\end{section}

\begin{section}{Conclusion}
\label{outro}
In this paper we presented the results of calculation of the
two-loop corrections to the relation between pole and running massess
of the $b$-quark, proportional to the Yukawa couplings of heavy quarks
($y_t, y_b$).  We provided a numerical analysis of the value of
these corrections in different regions of the CMSSM parameter space.
Corrections presented here can be used in a renormalization group
analysis of the Yukawa coupling unification.  It was also found
that $m_b$ corrections significantly affect low energy phenomenology
where the $b$-quark enters \cite{Ibrahim:2003jm}. The
analysis given in \cite{Gomez:2004ek} showed that the neutralino
relic density is very sensitive to the mass of the $b$-quark for
large $\tan\beta$, thus the result considered in this paper may have
important implications to the dark matter searches. We are going
to study these issues and calculate more ${\mathcal O}(\alpha_s^2)$
terms in the large mass expansion of the relation between the top pole
and $\overline{\rm DR}$ masses, which requires some improvement of
our code.

The authors would like to thank  M.Kalmykov, D.Kazakov,
A.Onishchenko, V.Velizhanin and O.Veretin for fruitful discussions
and multiple comments. Financial support from RFBR grant
\#~02-02-16889, the grant of Russian Ministry of Industry, Science
and Technologies \#~2339.2003.2, DFG grant 436~RUS~113/626/0-1 and
the Heisenberg-Landau Programme is kindly acknowledged.
\end{section}



\begin{thebibliography}{**}
\bibitem{Ellis:2004bx}
J.~R.~Ellis, K.~A.~Olive, Y.~Santoso and V.~C.~Spanos,
arXiv:hep-ph/0408118.

\bibitem{deBoer:2003xm}
W.~de Boer and C.~Sander,
Phys.\ Lett.\ B {\bf 585}, 276 (2004)
[arXiv:hep-ph/0307049].

\bibitem{Baer:2003wx}
H.~Baer, C.~Balazs, A.~Belyaev, T.~Krupovnickas and X.~Tata,
JHEP {\bf 0306}, 054 (2003)
[arXiv:hep-ph/0304303].

\bibitem{deBoer:2001xp}
W.~de Boer, M.~Huber, A.~V.~Gladyshev and D.~I.~Kazakov,
Eur.\ Phys.\ J.\ C {\bf 20}, 689 (2001)
[arXiv:hep-ph/0102163].

\bibitem{Abel:2000vs}
S.~Abel {\it et al.}  [SUGRA Working Group Collaboration],
arXiv:hep-ph/0003154.

\bibitem{Pierce:1996zz}
D.~M.~Pierce, J.~A.~Bagger, K.~T.~Matchev and R.~j.~Zhang,
Nucl.\ Phys.\ B {\bf 491}, 3 (1997)
[arXiv:hep-ph/9606211].

\bibitem{Bednyakov:2002sf}
A.~Bednyakov, A.~Onishchenko, V.~Velizhanin and O.~Veretin,
Eur.\ Phys.\ J.\ C {\bf 29}, 87 (2003)
[arXiv:hep-ph/0210258].

\bibitem{Siegel:1979wq}
W.~Siegel,
Phys.\ Lett.\ B {\bf 84}, 193 (1979).

\bibitem{Smirnov:2002pj}
V.~A.~Smirnov,
``Applied asymptotic expansions in momenta and masses,''
http://www.slac.stanford.edu/spires/find/hep/www?irn=4841620

\bibitem{Avdeev:1997sz}
L.~V.~Avdeev and M.~Y.~Kalmykov,
Nucl.\ Phys.\ B {\bf 502}, 419 (1997)
[arXiv:hep-ph/9701308].

\bibitem{Davydychev:1992mt}
A.~I.~Davydychev and J.~B.~Tausk,
Nucl.\ Phys.\ B {\bf 397}, 123 (1993).

\bibitem{Chetyrkin:1981qh}
K.~G.~Chetyrkin and F.~V.~Tkachov,
Nucl.\ Phys.\ B {\bf 192}, 159 (1981).

\bibitem{Hahn:2001rv}
T.~Hahn and C.~Schappacher,
Comput.\ Phys.\ Commun.\  {\bf 143}, 54 (2002)
[arXiv:hep-ph/0105349].

\bibitem{Bauer:2000cp}
C.~Bauer, A.~Frink and R.~Kreckel,
arXiv:cs.sc/0004015.

\bibitem{twoloopvargs}
A.~Sheplyakov, ``TwoLoop, C++ library for large mass expansion of 2-loop propagator-type diagrams''
(in preparation).

\bibitem{Fleischer:1999tu}
J.~Fleischer and M.~Y.~Kalmykov,
Comput.\ Phys.\ Commun.\  {\bf 128}, 531 (2000)
[arXiv:hep-ph/9907431].


\bibitem{Ibanez:1984vq}
L.~E.~Ibanez, C.~Lopez and C.~Munoz,
Nucl.\ Phys.\ B {\bf 256}, 218 (1985).

\bibitem{modelline}
S.P.~Martin, 
http://zippy.physics.niu.edu/modellines.html

\bibitem{Martin:2002zb}
S.~P.~Martin, S.~Moretti, J.~m.~Qian and G.~W.~Wilson,
in {\it Proc. of the APS/DPF/DPB Summer Study on the Future of Particle Physics (Snowmass 2001) } ed. N.~Graf,
eConf {\bf C010630}, P346 (2001).

\bibitem{Allanach:2001kg}
B.~C.~Allanach,
Comput.\ Phys.\ Commun.\  {\bf 143}, 305 (2002)
[arXiv:hep-ph/0104145].

\bibitem{Ibrahim:2003jm}
T.~Ibrahim and P.~Nath,
Phys.\ Rev.\ D {\bf 68}, 015008 (2003)
[arXiv:hep-ph/0305201].

\bibitem{Gomez:2004ek}
M.~E.~Gomez, T.~Ibrahim, P.~Nath and S.~Skadhauge,
Phys.\ Rev.\ D {\bf 70}, 035014 (2004)
[arXiv:hep-ph/0404025].
\end{thebibliography}
\end{document}